\documentstyle[epsfig]{mlaa}

\begin{document}
\title{The use of Petersen diagrams and period ratios\\ in
investigating the pulsational content \\of stars in the classical Instability Strip
}
\author{Ennio Poretti and Marco Beltrame}

\institute {INAF - Osservatorio Astronomico di Brera, Via E. Bianchi, 46 -- 23807 Merate (LC),
Italy}
\date{Invited review presented at the MinySymposium ``Asteroseeismology and Stellar Evolution",
JENAM, Budapest, 25-30 August 2003 -- To be published in {\it Communications in Asteroseismology}
Series, 2004} 
\maketitle
\begin{abstract}
The aspects of the analysis of  photometric time--series obtained on double--mode or
multiperiodic pulsating stars are briefly reviewed. In particular, the ratios between frequencies
are used to pin cases revealing peculiarities. In addition to the Petersen diagrams, 
we also demonstrated that the period ratios can detect interesting objects. 
In particular, new results are obtained on High--Amplitude Delta Scuti contained in the
OGLE-II database.
\end{abstract}
\section{Introduction}
In the recent years a huge collection of time--series has been obtained on variable
stars, as a noticeable by--product of several microlensing projects. Therefore, the
investigation of thousands of light curves is carried out by detecting the frequencies
present in the time--series. It is not easy to give astrophysical depth to this kind of analysis, as
we have monochromatic photometric data only at our disposal.
Different classes of variable stars can show very different physical processes
with very similar light curves.  
Moreover, instrumental terms are often superimposed to physical ones in the
same frequency range.
In this paper we will try to review
some of these aspects.

\section{The Petersen diagrams}
When multiperiodicity is detected, it is quite an obvious procedure
to calculate the frequency
ratios. In case of double--mode pulsation, we deal with a short ($P_S$) and a
long ($P_L$) period. The plot of $P_L$ versus the 
$P_S/P_L$ ratio is now known as the Petersen diagram (Petersen 1973). 
Small 
variations of temperature, mass, metallicity or luminosity push the
$P_L$, $P_S/P_L$ point into different
directions: as an example, see Fig.~2 in Popielski et al. (2001). 
The mass capable to reproduce the observed $P_S/P_L$ ratio is called 
the beat mass. 

There have been several astrophysical applications of the
Petersen diagrams. The most relevant one has been the reconciliation of
pulsational, evolutionary and beat masses of Double Mode Cepheids
after having introduced the 
new opacities (Christensen-Dalsgaard \& Petersen 1995). 
Let us indicate the fundamental radial mode as $F$, the first overtone
radial mode as 1O, the second one as 2O.

\begin{figure}
\centering
\resizebox{\hsize}{!}{\includegraphics{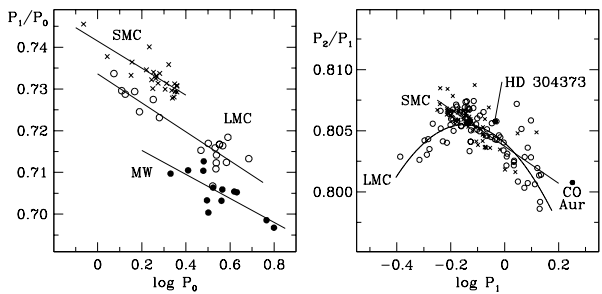}}
\caption{The Petersen diagrams for double--mode Cepheids
in the Milky Way, the Small and Large Magellanic Clouds. The $P_0/P_1$
pulsators 
belonging to the three different galaxies are well separated, while
the $P_2/P_1$ ones are not.
}
\label{dmc}
\end{figure}

Recently, Beltrame \& Poretti (2002) demonstrated that HD 304373 is the second case of 
1O/2O double--mode Cepheid in the Galaxy. Its position in the Petersen diagram
is very similar to that of some LMC stars. 
In general, all the 2O/1O values are very close to each other and only the
different range of periods allows us to separate the different environments
(Fig.~\ref{dmc}, left panel). On the contrary, the 1O/F ratios show a strong dependence on
metallicity (Fig.~\ref{dmc}, right panel).
It seems that
the 1O/2O ratios are less sensitive than the 1O/F ratios to the difference in metallicity:
the latter values are ranging on a five time larger interval than the former ones. 

Christensen-Dalsgaard \& Petersen (1995) pointed out
that the matching between the F/1O ratios for galactic pulsators
and the theoretical models occurs for metallicities smaller than
the solar value of 0.017--0.020; if that applies for the 1O/2O
pulsators too, a metallicity close to 0.010 allows the $P_2/P_1=0.8058$
and $P_1$=0.922405~d values to reasonably fit the theoretical
models (Christensen-Dalsgaard \& Petersen 1995).
We also note that assuming $P_1/P_0$=0.70, we obtain $P_0$=1.32~d,
i.e., the fundamental period of HD~304373 is really one of the shortest
among galactic Cepheids.

As in the case of CO Aur, there is no significant contribution of the 2$f_2$,
harmonic in the light curve of HD~304373, i.e., it is perfectly sine--shaped
within the error bars. 
Only small or marginal deviations from a sine wave have been found in the LMC
and SMC 2O samples.

\section{The distinction between pulsating stars and geometrical variables}
The analysis of the 82 periodic pulsating stars included in the OGLE II database
(Mizerski \& Beijger 2002) revealed that pulsating and geometrical variables can be confused by some 
automatic routines, but they can be more easily separated by more clever 
methods (Beltrame 2002).  As a
first step, we decided to analyse in frequency all the stars,
to avoid any possible misinterpretation.
Using the least-squares iterative sine wave fitting (Vani\v{c}ek 1971), 
we obtained the power spectrum for every variable star
of our sample, thus being able to compare the highest peaks with their 
aliases. No discrepancy was found as respects the values reported by
Mizerski \& Beijger (2002).
\newline
The following step was to refine the periods of all the stars and to analyze their residual power
spectrum, looking for the presence of multiperiodicity or other 
peculiar effects (such as the Blazhko effect for RRab and RRc). Therefore, we considered 
the preliminary solutions and the fit order found in the previous step. The 
solution was given as input parameters to a code keeping locked the relationships between
the main frequency and the harmonics terms (Multiple TRigonometric Analysis
Program; Carpino et al. 1987); 
the best fit was looked for around the preliminary solution. Once the refined 
period was obtained, we started searching for additional
terms. Only the frequency values of the main term and its harmonics were 
considered as input values; no prewhitening was done. A few multiperiodic
stars (RRab and RRc showing Blazhko stars, a new multimode pulsator) have
been discovered.
Moreover, the results of the period refinement shows that we have $\Delta P \sim 10^{-7}$ for 14 stars,
$\Delta P \sim 10^{-6}$ for 43 stars. The rest of the sample (24 stars) 
showed a more significant refinement, with $\Delta P\sim 10^{-5}$--$10^{-4}$.


\begin{figure}
\centering
\resizebox{\hsize}{!}{\includegraphics{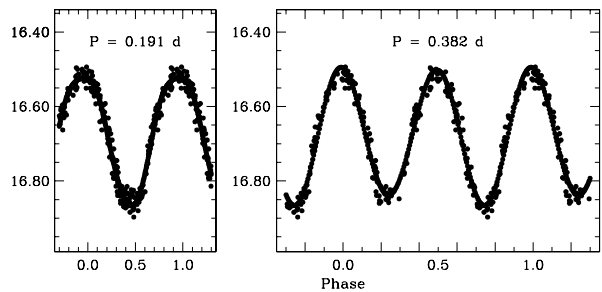}}
\caption{
The true nature of bul1.686: the star is not a pulsating variable with a
period of 0.191~d (left panel), but
a geometrical one with a period twice the ``pulsational" one (right panel): note the different depths
of minima.}
\label{wuma}
\end{figure}

Having obtained the correct periods for every star, we started with a preliminary fit  up to the
eighth order, in order to obtain the correct Fourier parameters. Therefore, we discarded every
harmonics with amplitude less than 3 times their error. During this procedure,  three stars
(bul1.2457, bul1.1323, bul1.686), classified as HADS by Mizerski \& Beijger (2002), showed a notable
spreading of points at the minimum light of their light curves. 
Therefore, we checked the possibility for them being
W~Uma variables, doubling the period found. We found that bul1.686 is
clearly a W UMa variable, since the light curve on the doubled period shows two
minima having  different amplitudes (Fig.~\ref{wuma}). In the other cases,
there is still some dispersion at minima and a sure classification cannot be made.
However, the power spectra always detected a couple of terms, in the ratio 2:1,
with the highest amplitude related to the shortest period.
A conservative approach would strongly suggest that in presence
of the 2:1 ratio the variable should be considered a geometrical one (eclipsing or
rotational). Indeed, to explain such a ratio by the excitation of resonant modes 
seems an ``ad hoc" solution, even if very attractive  for theoreticians.
We also note that stars showing a sine--shaped light curve (i.e., no significant
contribution from the harmonics of the main frequency) can be geometrical variables having
minima of equal amplitudes. In case of monoperiodicity, this fact makes it quite delicate to 
distinguish the sine--shaped
light curves of 2O pulsators (see previous section) from those of geometrical 
variables.
\section{The frequency ratios in High Amplitude Delta Sct stars}
\begin{figure}
\centering
\resizebox{\hsize}{!}{\includegraphics{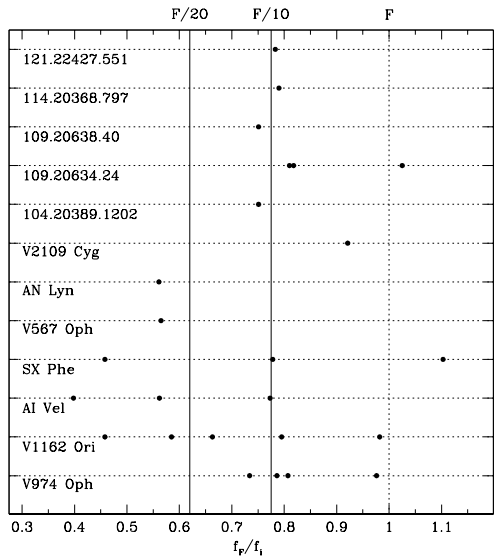}}
\caption{
Observed (filled dots) frequency ratios among HADS stars
showing possible nonradial modes. Vertical lines indicate the theoretical
ratio between fundamental (F) and overtone (first, 1O; second, 2O) radial
modes. Note also the ratios around 1.0, i.e., the presence of modes close to
the fundamental radial one.}
\label{ratio}
\end{figure}

Poretti (2003) recently pointed out how HADS are showing a frequency spectrum much more
complicated than the one considered some years ago.
Figure~\ref{ratio} (which is an updating of Fig.~5 in Poretti 2003)
illustrates  how several stars deviate from the canonical 0.77 ratio
between the $F-$ and the 1O modes. Values higher than 0.77 can be originated by
very low metallicities ($Z<$0.0005). Since most cases are actually much higher
than 0.77, the metal shortage should be very relevant and this seems unlikely for
Pop.~I stars.
Also fast
rotation can strongly modify the ratio, but it is quite uncommon in HADS stars, which
are evolved ones. 
If we also consider the variety of ratios visible in Fig.~\ref{ratio}, the excitation of 
nonradial modes can be indicated as  the natural explanation 
for such a different pulsational content. To corroborate the excitation of
nonradial modes in HADS, high--resolution spectroscopy is strongly recommended
to detect the signature of line profile variations.

\begin{figure}
\centering
\resizebox{\hsize}{!}{\includegraphics{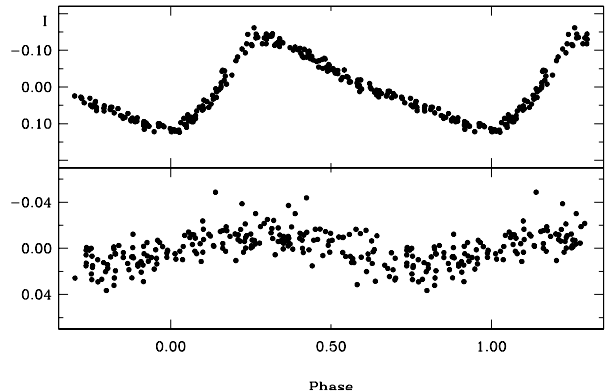}}
\caption{
The light curves of two frequencies detected in
the bul1.3074 time--series: they give $f_1/f_2$=0.62}
\label{bul}
\end{figure}

As discussed above, 
the 0.800 ratio is found in double mode Cepheids pulsating
in the 1O and 2O modes.
Such a value is also observed in some double--mode HADS stars (Musazzi et al. 1998)
and it is attributed to the same modes, but there are some unclear points.
In particular, the 0.800 ratio is related to the unusual shape of the
$P_L$ light curve: the descending branch is steeper than the ascending
one. This feature still remains unique in the zoo of the light curves of pulsating
variables.
The large databases did not supply any new stars simultaneously showing the 
0.800 ratio and the asymmetrical--in--the--bad--sense light curve. 

When looking at frequency ratios it is interesting to search for the 0.620 value,
which should be the signature of the excitation of the $F$ and  2O modes.
Performing the search described above in the OGLE-II database, we discovered that
bul1.3074 is a new multiperiodic
HADS star. Three independent frequencies 
have been detected, i.e., $f_1$=5.275, $f_2$=8.672 and $f_3$=8.618~cd$^{-1}$. In addition
to the close doublet composed of $f_2$ and $f_3$,
the ratio $f_1/f_2$ is intriguing, as it results in a 0.61 value. 
The presence of aliases could have interfered in determining the right frequencies, but even
considering the peaks at $f+1$, $f-1$~cd$^{-1}$  we were not able to solve the problem
detecting the usual 0.77 ratio. Also in this case a possible explanation could be the 
excitation of a nonradial mode of pulsation.
However, if  F/2O pulsators do exist among HADS, 
bul1.3074  should be considered as a promising candidate. Figure~\ref{bul} shows
the light curves of the $f_1$ and $f_2$ terms: they look very well defined, strongly
supporting their reality.

\section{Acknowledgments}
This paper was partly prepared during the EP's stay at Konkoly Observatory in the framework
of the Italian--Hungarian T\'eT cooperation (project I--24/1999).

\section {References}
Beltrame, M., 2002, Laurea Thesis, Universit\`a di Milano (in Italian)\\
Beltrame, M., Poretti, E., 2002, A\&A 286, L9\\
Carpino, M., Milani, A., Nobili, A.M. 1987, A\&A 181, 182\\
Christensen-Dalsgaard, J., Petersen, J.O., 1995, A\&A 229, L17\\
Mizerski, T., Beijger, M., 2002, Acta Astron. 52, 61\\
Musazzi, F., Poretti, E., Covino, S., Arellano Ferro, A., 1998, PASP 110, 1156\\
Petersen, J.O., 1973, A\&A 27, 89\\
Popielski, B. L., Dziembowski, W. A., Cassisi, S., 2000, Acta Astron. 50, 491 \\
Poretti, E., 2003, A\&A 409, 1031\\
Vani\v{c}ek, P. 1971, Ap\&SS 12, 10\\
\end{document}